# Fundamental longitudinal electromagnetic (EM) force investigation using DC current

Neal Graneau
AWE Nuclear Security Technologies

# Fundamental longitudinal electromagnetic (EM) force investigation using DC current

Neal Graneau

*Abstract*— **The purpose of this work was to investigate historical claims of the existence of a longitudinal ElectroMagnetic (EM) force component acting on metallic atomic current elements in a direction parallel to the current flowing through them. This lies outside conventional textbook physics predictions, yet its reality has been indicated previously and if eventually confirmed will have a significant effect on physics theory and many technological applications, especially those involving high current density ( > $10^9$ A/m$^2$) including EM mass drivers such as railguns with liquid metal sliding contacts, flux compression generators and also high energy plasmas including fusion and other energy technologies. The experiment described here is based on the measurement of force on a supported copper armature submerged in a trough containing liquid metal through which constant DC current is passing. The coaxial symmetry of the experiment (CRE-Coaxial Recoil Experiment) was able to limit all net force on the centred armature to the direction of interest, parallel to the current within it. All the experimental data related the measured forces on the armature to the total current flowing through the circuit. Variations were made to the length and location of the armature along the central axis as well as modifications to the liquid metal configuration. Raw and processed results are presented, and an experimental technique is described that revealed strong evidence for the existence of, and discrimination between, axial mechanical contact forces between solid and liquid metal as well as longitudinal EM force. The original EM force law, proposed in 1822 by Ampère, includes a longitudinal component and has been found to be qualitatively consistent with all experiments to date, including these reported findings, and is considered a candidate explanatory theory.**

## I. Longitudinal EM Force

The possible existence of a longitudinal (parallel to the current) electromagnetic EM force on a macroscopic electric conductor has been debated and researched for more than 200 years. Historically, the adoption of Maxwellian field theory and the associated Lorentz force law ($\mathbf{F}_L = \mathbf{J} \times \mathbf{B}$) only allows for the prediction of a transverse force density, $\mathbf{F}_L$, perpendicular to the current density, $\mathbf{J}$, by the mathematical definition of the vector cross product. However, due to the publication of sporadic experiments over the last 200 years [1][2][3][4][5][6][7], the existence of a longitudinal EM force component has been supported and never discounted, although certainly never accepted as fact.

Neal Graneau is with AWE Nuclear Security Technologies, UK (e-mail:neal.graneau@awe.co.uk).

If longitudinal EM forces exist, they are expected to be most significant in high current pulsed power applications with current density exceeding $10^9$ A/m$^2$, including railguns [8], exploding wires [2][9], anomalous forces in water electrospray and water-arc explosions [10][11], vacuum arc cathode spots [12], Z-pinches [13], Liner Implosions [14], Plasma Focus [15], Flux Compression Generators [16], Flyer Plate acceleration [17] and probably others. At lower current densities, MHD liquid metal pumping along electric current streamlines, without the use of external magnets, is possible and has been observed to occur in liquid metal devices such as furnaces and batteries whether desired [18][19] or not [20]. It is also expected to be possible to observe longitudinal EM force acting on solid conductors within liquid metal pools [1][2][3] which is the motivation for the present experiment described here (CRE – Coaxial Recoil Experiment). Previously, several investigations into the existence of longitudinal EM force in pulsed power circuits have been ambiguous due to the transient and explosive nature of the experiments [2][4][5][6][9][12], thus leading to the motivation to perform the recent research presented here with low density constant DC current acting for more than 10 seconds, leading to steady force measurements.

To avoid ambiguity, there is a historical linguistic confusion in this subject area. EMF (ElectroMotive Force) is not a force but relates to potential energy gained or lost by charged particles due to inductive effects, caused by changing current amplitudes and/or conductor geometries, and is measured in Volts (V). In contrast, the research discussed here investigates EM (ElectroMagnetic) Force, which is contrastingly described as mechanical, can accelerate free charged particles or neutral current carrying matter, is measured in Newtons (N) and can be caused by DC as well as varying currents and both static and mobile circuits.

## II. CRE – Coaxial Recoil Experiment

### A. Experimental Setup

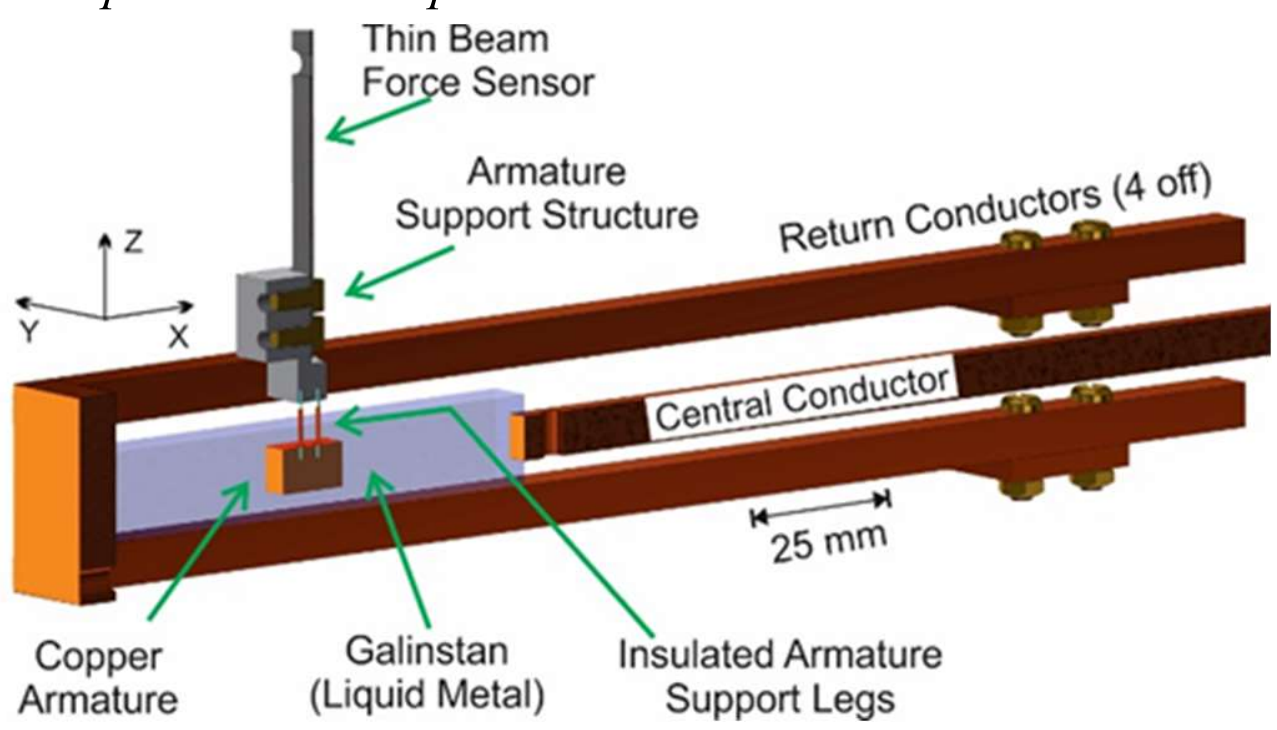

**Fig. 1.** Centre section schematic of the CRE. Liquid metal trough containment not shown for clarity.

The CRE was deliberately designed to examine the possible longitudinal EM recoil forces which have been previously observed in railgun-like circuits [2][6]. It exploited a coaxial geometry so that all objects on the central axis would experience zero net transverse EM and mechanical forces by symmetry. A section view of the experiment is depicted in Fig. 1. The longitudinal axis of the central copper conductor and liquid metal trough defined the X-axis, which was surrounded by four symmetrically positioned copper return conductors, all parallel to the X-axis. The return conductors were electrically connected to each other at the corners of a fixed 28 mm x 28 mm x 6 mm copper plate at one end of the trough. All five long fixed copper conductors were square cross section (6 mm x 6 mm) with their lateral surfaces parallel to the X-Y or X-Z planes. The liquid metal was Galinstan, a conductive, non-toxic eutectic alloy of Gallium, Indium and Tin which is liquid at room temperature and had a square cross section of nominally 12 mm x 12 mm (actual width = 11.9 mm, depth of fill = 12 ± 0.5 mm). A K-type thermocouple was inserted into the dielectric trough wall within 1 mm of the liquid metal. The armature was a copper cuboid with two opposing 6 mm x 6 mm exposed end faces , parallel to the Y-Z plane, and insulated on its other four lateral surfaces with a composite of insulative varnish and Kapton tape. The armature was held in position, centred on the X-axis in the liquid metal trough, by a beam force sensor (Strain Measurement Devices SMD100-0002 200 mN) as shown in Fig. 1. The exposed armature end faces were perpendicular to the central X-axis, and it was connected to the force sensor by an insulated structure. The sensor was orientated specifically to measure force parallel to the X-axis, however the force on the armature was not acting directly on the calibrated force application location on the sensor beam. As a consequence, the lever arm had been extended by a factor of 1.567. Therefore, all force measurements recorded with the CRE were the sensor calibrated readings multiplied by 0.638 (= 1/1.567). The corrected rated deflection of the sensor was then approximately 27 μm/mN, and the largest measured forces were of the order of 2 mN, with a consequent maximum armature displacement of ~54 μm from its neutral (zero current) position. Therefore, for the accuracy of armature location in the CRE (± 0.1 mm), it does not move significantly during a current pulse. The location of the armature could be precisely controlled by a 3-axis XYZ micrometer driven translation stage which set the location and orientation of an aluminium sensor mounting block to which the fixed end of the sensor beam was firmly attached, as indicated in Fig. 2. The experimental circuit was connected to a low voltage 355 A output DC constant current power supply. Figure 2 displays the current streamlines for a particular polarity of connection to the power supply. However, all of the results reported here were found to be independent of the power supply polarity. Due to the difference in resistivity between copper and Galinstan, it is estimated that 95% of the total current flows through the copper armature with the remaining 5% passing uniformly around the armature in the liquid metal.

Fig. 3 depicts the dimensions of the CRE. All lengths are expressed in mm. A is the length of the armature and three different ones (A = 6, 10 & 16) were deployed during the experiment. GA and GB are the lengths of the liquid metal column between either end of the armature and the fixed electrodes at the end of the nominally 70 mm long trough (actual length = 69.6 mm). A thin Kapton insulator was glued to the inner surface of the trough end plate that forced current to flow into the Galinstan only via the central 6 mm x 6 mm square area aligned to the armature and central electrode end faces. All copper parts were manufactured to ± 0.2 mm tolerance. All circuit dimensions are given for the purpose of future modelling attempts.

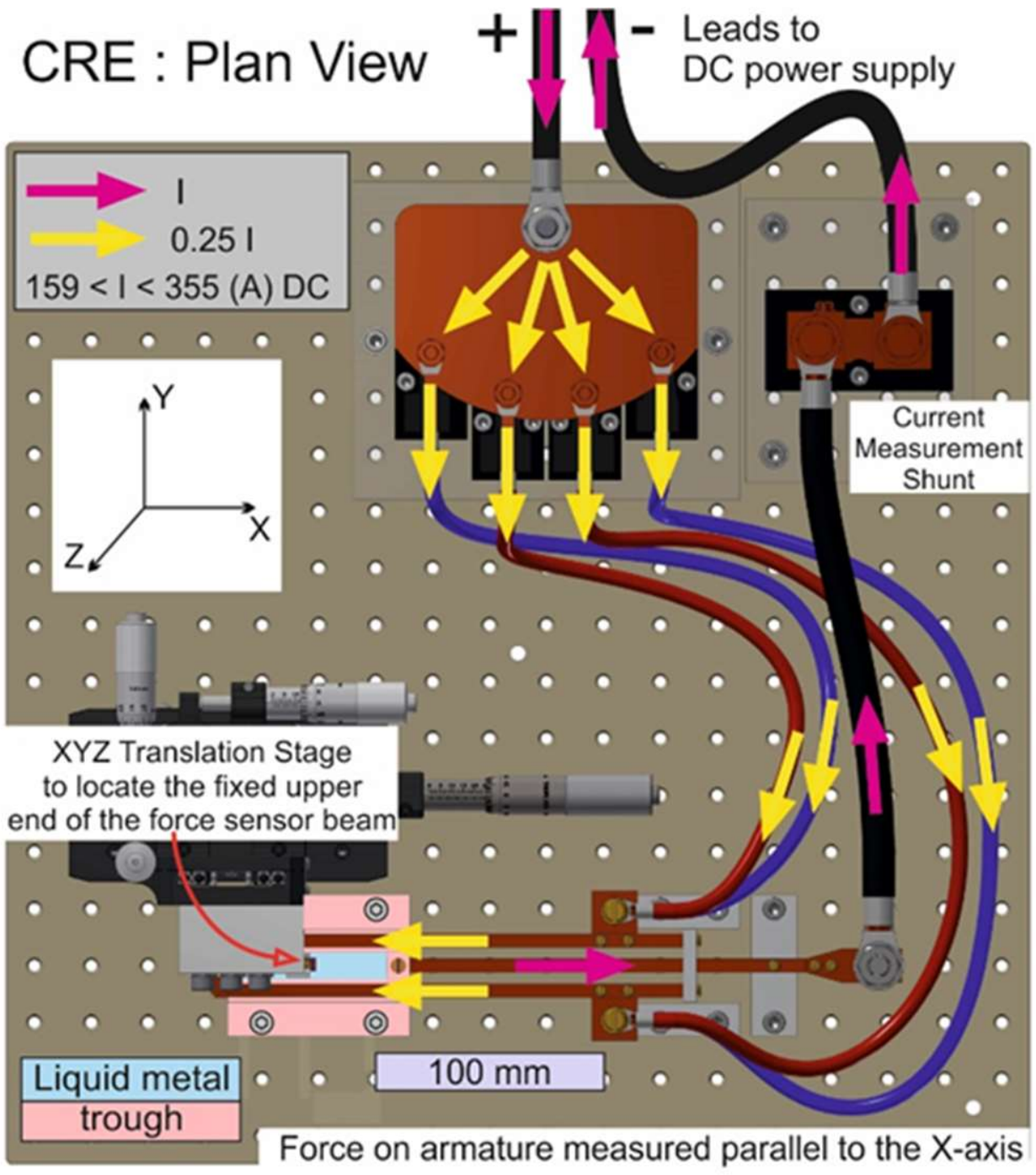

**Fig. 2.** Plan view of the main components of the CRE including arrows indicating the flow of current. The power supply can be attached with the opposite polarity, thus reversing all current directions with no change in experiment performance.

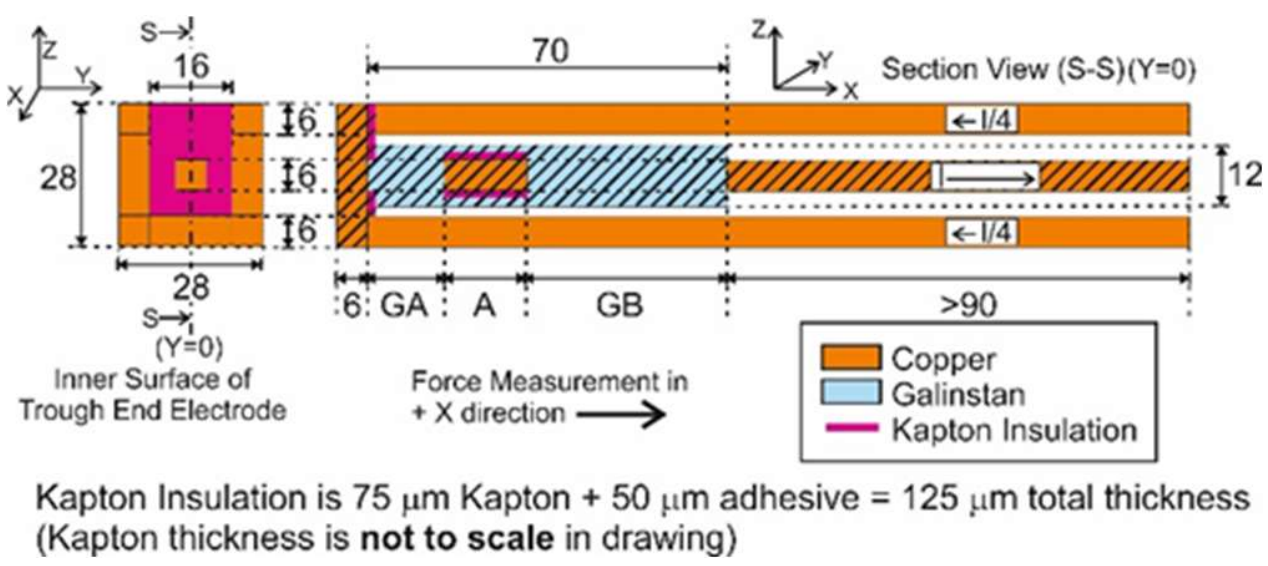

**Fig. 3.** Dimensions (mm) of all conductors in the CRE and diagrammatic definitions of the parameters A, GA and GB. Current direction and distribution in the axial and return conductors is also displayed. Dielectric trough and connecting screws have been removed for clarity.

*B. Raw Data*

For a selected armature length and location, 12 force measurements were taken using 6 differing constant DC current magnitudes over approximately 10 minutes. A typical set of

~15 s duration force and current pulses is shown in Fig. 4. While the measured force varies by a factor of five over this range of currents, it is always found to be approximately proportional to the circuit current squared for any given setup of A, GA and GB. As a result, a current independent force constant, K (as defined in Fig. 4) and standard deviation, $\sigma_K$, can be deduced for each armature and location. This proportionality indicates that the net armature force is likely to be caused by one or more force mechanisms of EM origin.

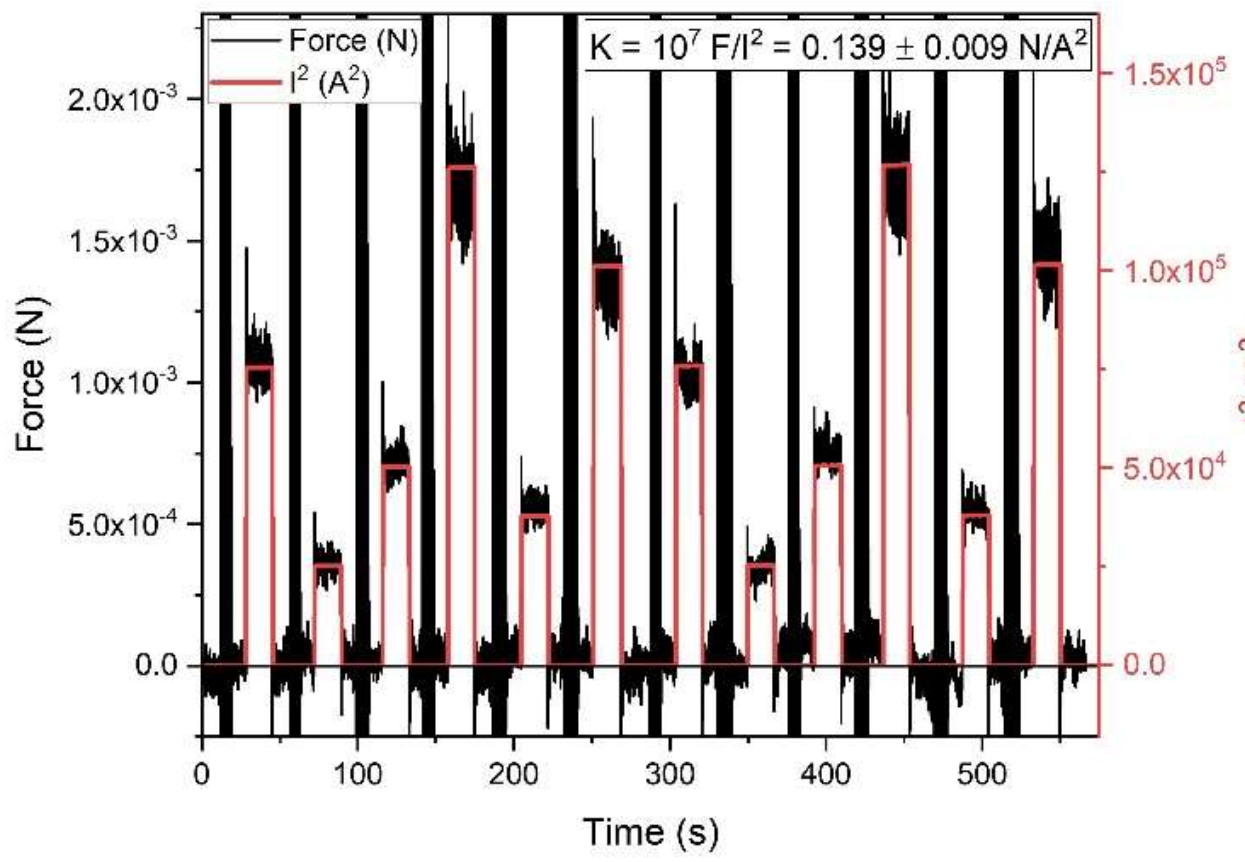

**Fig. 4.** Force and current measurement during a typical 12 current pulse sequence for A=10, GA=0.5, GB=59.11, K=0.139, $\sigma_K$=0.009, demonstrating that the force is proportional to the square of current and defining K as a current independent force constant.

The thick bands of measured force that occurred between the individual force and current pulses in Fig. 4 are caused by manually lightly tapping the sensor to allow the armature and support legs to more easily overcome the surface tension and viscosity of the Galinstan and reset to its neutral (Force ~ 0) location for pulse-to-pulse consistency. No monotonic effect on K due to the small, measured temperature rise (4-7°C) over the 12-pulse sequence could be detected.

As an example, all the data using the A = 10 mm armature is shown in Fig. 5. Each point on Fig. 5 represents a data sequence like the one shown in Fig. 4 and the red line is a best fit to indicate trends. The dot is the average K value, and the error bars represent ± $\sigma_K$. The average K values shown on the graph represent current independent force measurements that can be repeatedly correlated to the armature location, which can be parameterized by either a GA or GB value as defined in Fig. 3. In Fig. 5, various points are noted in the legend and will be referred to in the next sections. The observations common to all force measurements on all 3 armature lengths are that it experiences a force in the +X direction when GA is small and in the -X direction when GB is small, i.e., repulsion from the ends of the trough, and it also experiences only a very small (or zero) force when it is situated in the middle of the trough (GA ≈ GB). As GA and GB are increased from zero, the measured force magnitude initially decreases and in general it then reverses sign before eventually becoming approximately zero when in the middle. All the data for all 3 armature lengths is summarized in Fig. 6. While it is clear from Fig. 6 that armature length, A, plays an important role in determining the measured force constant, K, it is not easy to discern a clear numerical trend across the large ranges of GA or GB. Consequently, it is most informative to concentrate on the data when GA or GB is near or equal to zero (armature near or touching an end electrode) where it is noticeable that the force on the 6 mm armature is less than for the longer ones. However, the measured values of the net force on the two longer armatures are within each other's standard deviations and are thus not clearly distinguishable from each other.

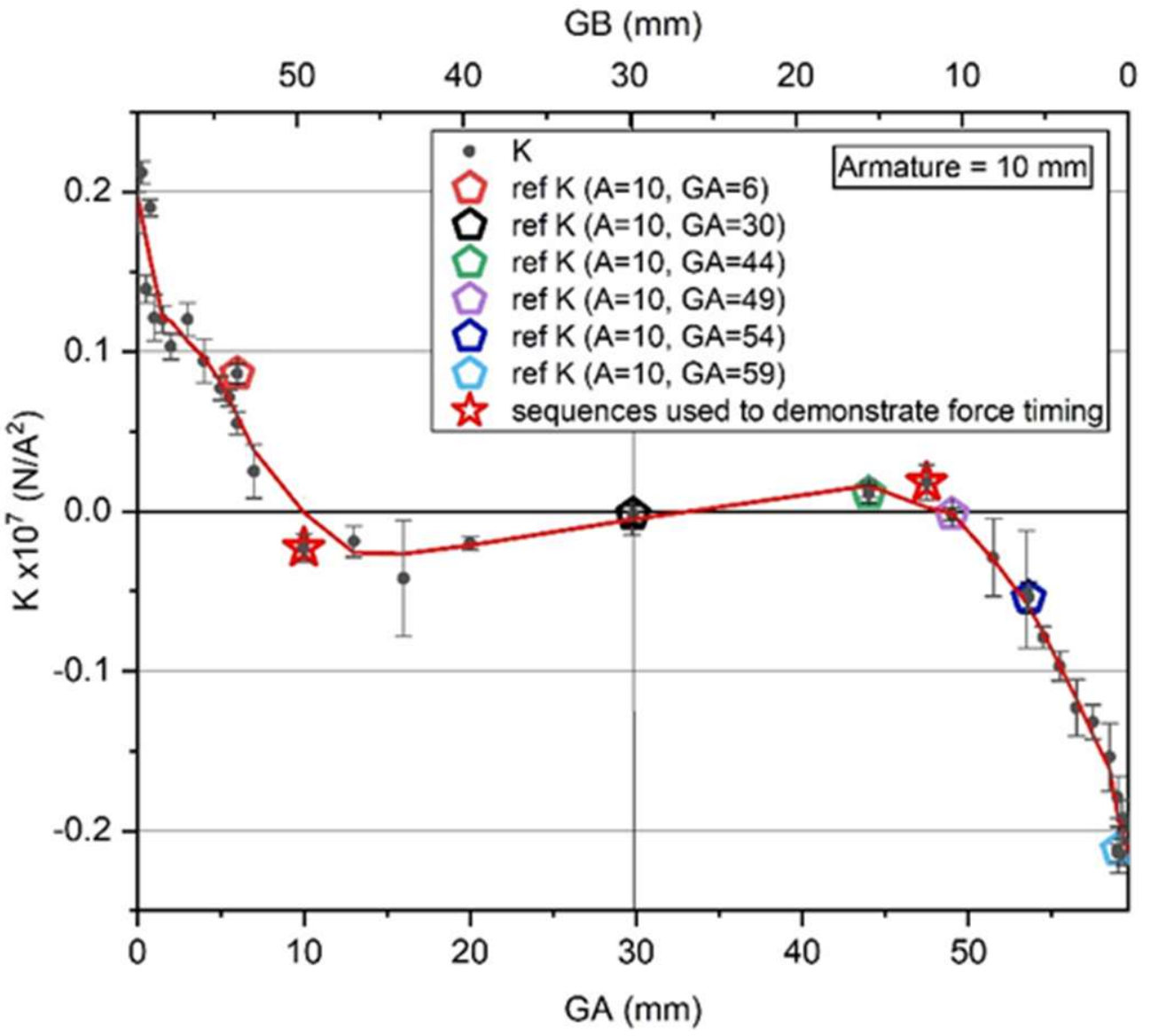

Fig. 5. The current independent force constant, K, for all measurements with armature A = 10 mm.

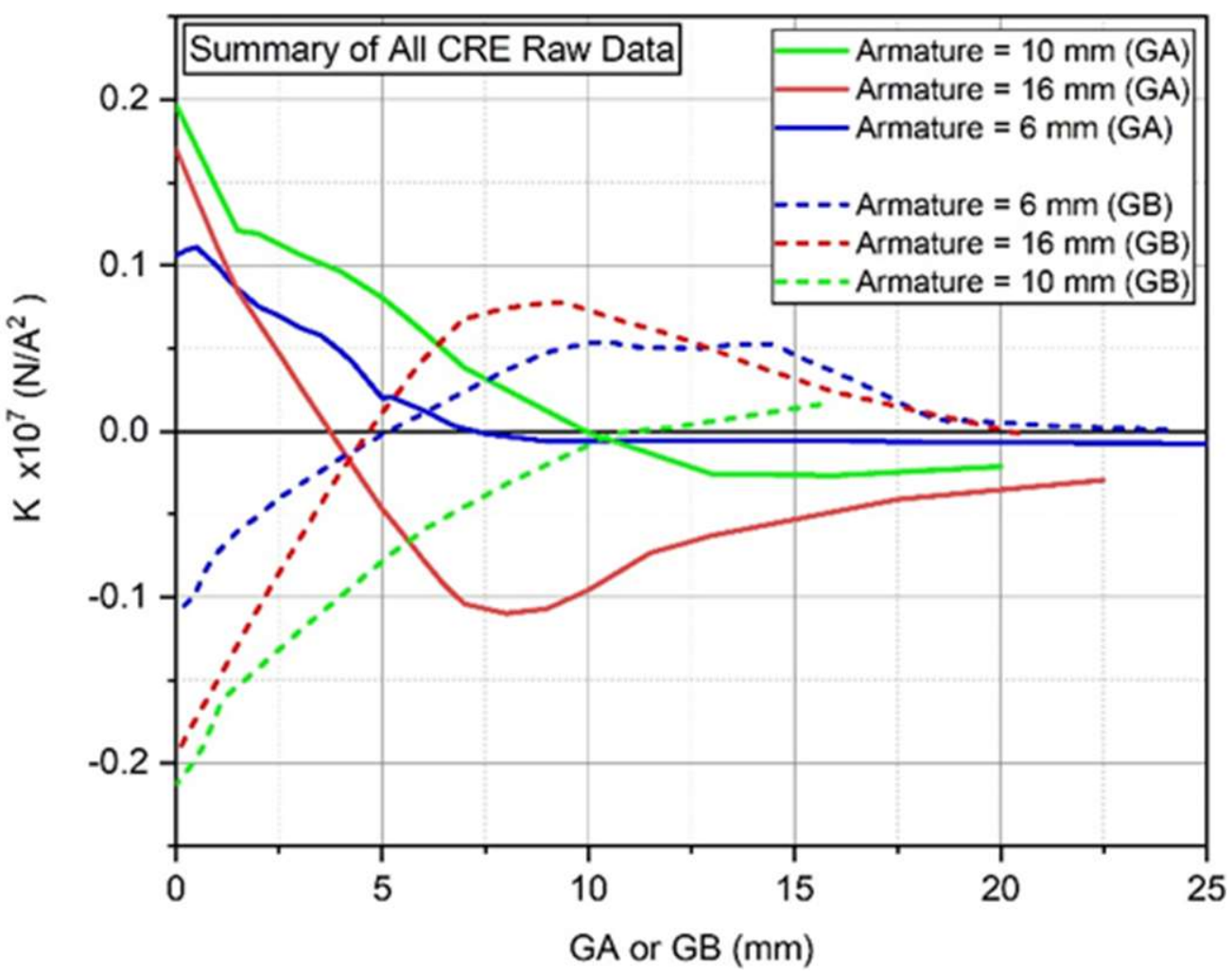

**Fig. 6.** A summary of the measured current independent force constant, K, for all 3 armatures in all locations.

It is presumed that the complexity of the data in Fig. 6 stems from the possibility that the armature is simultaneously subject to EM forces as well as to mechanical hydraulic forces caused by contact with the liquid metal. Without a full MHD model of the system, it is simply not possible to even attempt to quantitatively predict this data pattern, but it is reported here in the hope that such modelling and comparison may be performed in the future.

It had been hoped to confirm by measurement that there was zero net transverse force on all armatures at all locations as expected by the circuit's coaxial symmetry. However, there was no time to perform these measurements. For completeness, however, it was confirmed that all reported results were independent of the direction of current flow, a further indication that all forces were a function of the square of the circuit current and thus likely to be of EM origins.

### *C. Distinguishing between Axial EM and Mechanical Forces*

Regarding the different force mechanisms acting on the armature in the CRE, the direct contact hydraulic pushing forces caused by the movement of liquid metal are considered to be **MECH**anical whereas the **EM** force is purely a consequence of currents flowing in all conductors in the circuit. Both forces can be represented by current independent net force factors, $K_{MECH}$ and $K_{EM}$ respectively, which must sum to the net measured force constant, K. Therefore, the CRE could only reveal evidence of the existence and magnitude of an axial EM force if the mechanical force could be independently estimated ($K_{EM} = K - K_{MECH}$).

The possible mechanisms that can lead to a mechanical force can be either (a) independent of the length of the liquid metal column creating it or (b) dependent on the column length. In case (a), the forces would be equal and opposite on both ends of the armature wherever it is located and therefore always apply zero net force. However, if there is a length dependent mechanism causing a net mechanical pushing force on the armature, then a technique was developed to attempt to measure it.

The most studied EM effect on a liquid metal current carrying conductor was described in 1907 by Hering [21] as the "Pinch Phenomenon" and first observed in channels in metal smelting furnaces. In the same year, his colleague, Northrup [22] was the first to measure and calculate the transverse pinch in axial current carrying liquid metal troughs and examine the consequent axial mechanical pressure acting on the electrode faces at the ends of the channel. Based on the EM force principle discovered by Ampère [1], of the attraction of parallel current filaments, he derived that the resulting mechanical pressure on the ends of the column exists, but most importantly, does not depend on the channel length. The EM pinch phenomenon was later more widely studied by Bennett [23], but by the 1930's, it was mainly applied to dense plasmas including fusion devices. It has recently been applied to the development of LMCL (Liquid Metal Current Limiters) [24][25]. The conclusion here is that if a length dependent axial mechanical force exists in the CRE liquid metal, then it is not explained by the transverse force derived pinch mechanism, but must have another cause, either thermal or possibly longitudinal EM force acting on and within the liquid.

The measurement of a net axial, length dependent, mechanical force acting on the CRE armature was successfully achieved by placing a stiff but thin (100 μm) stainless-steel barrier at certain locations in the trough which completely filled its cross-section and then measuring the difference in force between before and after this modification. It was assumed that the barrier was stiff enough to prohibit the transfer of mechanical force from one side to the other and yet would not significantly distort the current streamlines as long as it was at least 4 mm from either an end of the armature or a fixed electrode at an end of the trough. The barrier is shown in Fig. 7 on its own and inserted in the dielectric trough. The length scale on the trough was used to determine its location. The force sensor beam, the top of the armature support structure and the thermocouple are also highlighted.

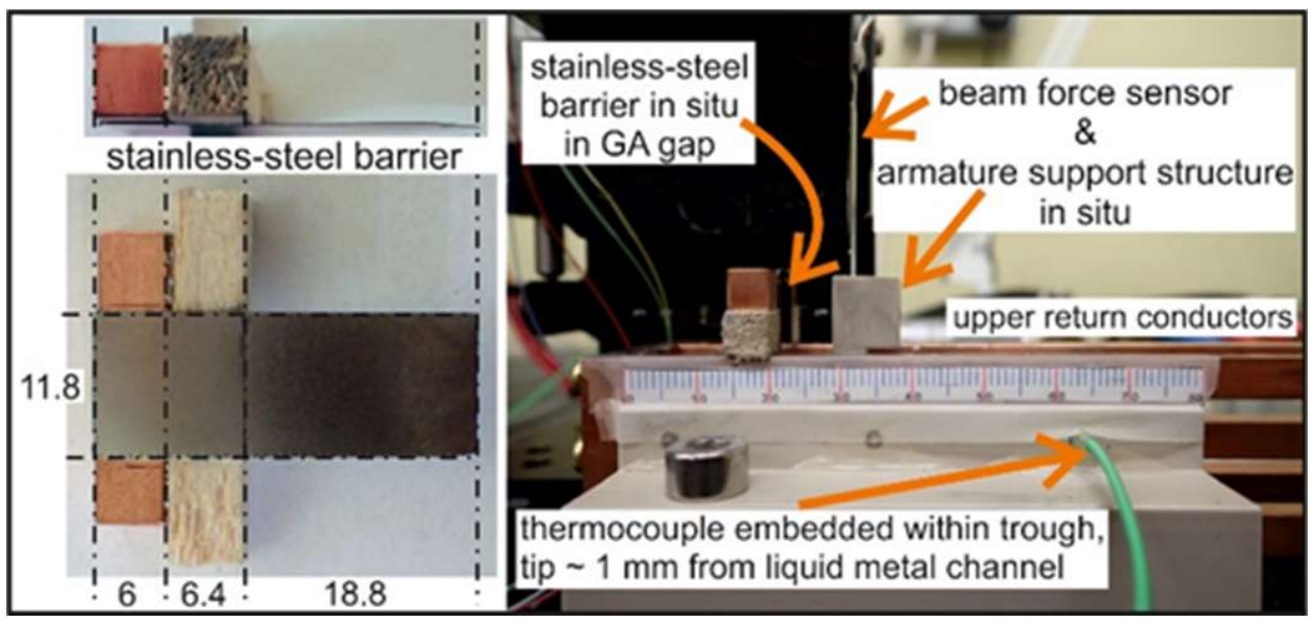

**Fig. 7.** Views of stainless-steel barrier, beam force sensor and support structures, dielectric trough and thermocouple.

The barrier therefore reduced the effective length of either GA or GB and created a new length parameter, LR, as defined in Fig. 8. Thus, if there is a length dependent mechanism that causes a column of liquid metal to expand along its length parallel to the current, then this technique should reveal it and ultimately yield a method of estimating the mechanical force applied by any length of this column.

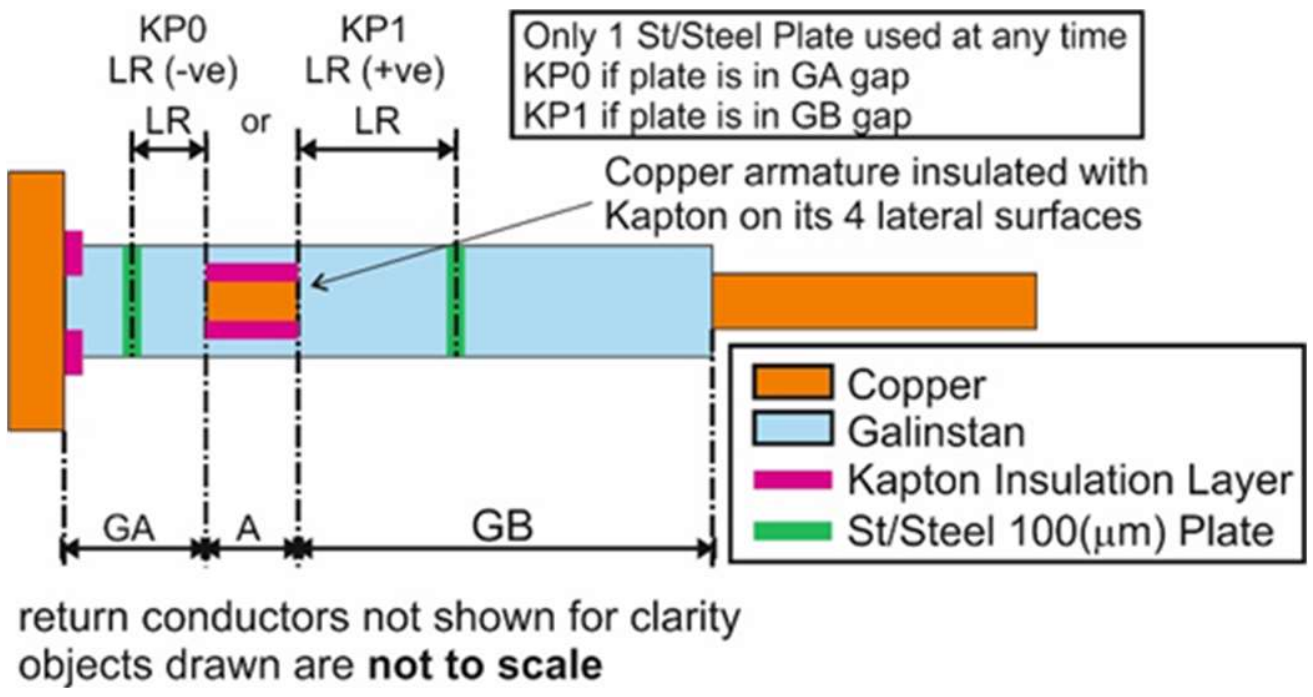

**Fig. 8.** Central section of Galinstan trough, yielding definitions of the parameters LR, KP0 and KP1 used in the method to estimate the mechanical force on the armature.

The technique used as references, six of the data points from Fig. 5 with A = 10 mm, noted on the legend as related to a particular barrier data series, each defined by an A and GA value. There were also two further reference points taken from the A = 6 mm data (not shown in this paper). Thus, there were eight sets of generated data, each with a specific A and GA value and a measured reference, K(A,GA). Force measurements were then taken by placing the barrier at multiple locations in the GA or GB column and determining the length and sign, LR, which was taken to be negative if in the GA region (KP0) and positive if in GB (KP1). 55 of these data sets were taken in this manner spread amongst the eight reference point conditions and yielded a **M**odified force constant which was now a function of LR in addition to A and GA, and called $K_M$(A,GA,LR). This allowed the effect of the barrier modification to be quantified, $\Delta K_M$, as shown in Fig. 9, yielding two apparently straight-line relations for the KP0 and KP1 data respectively.

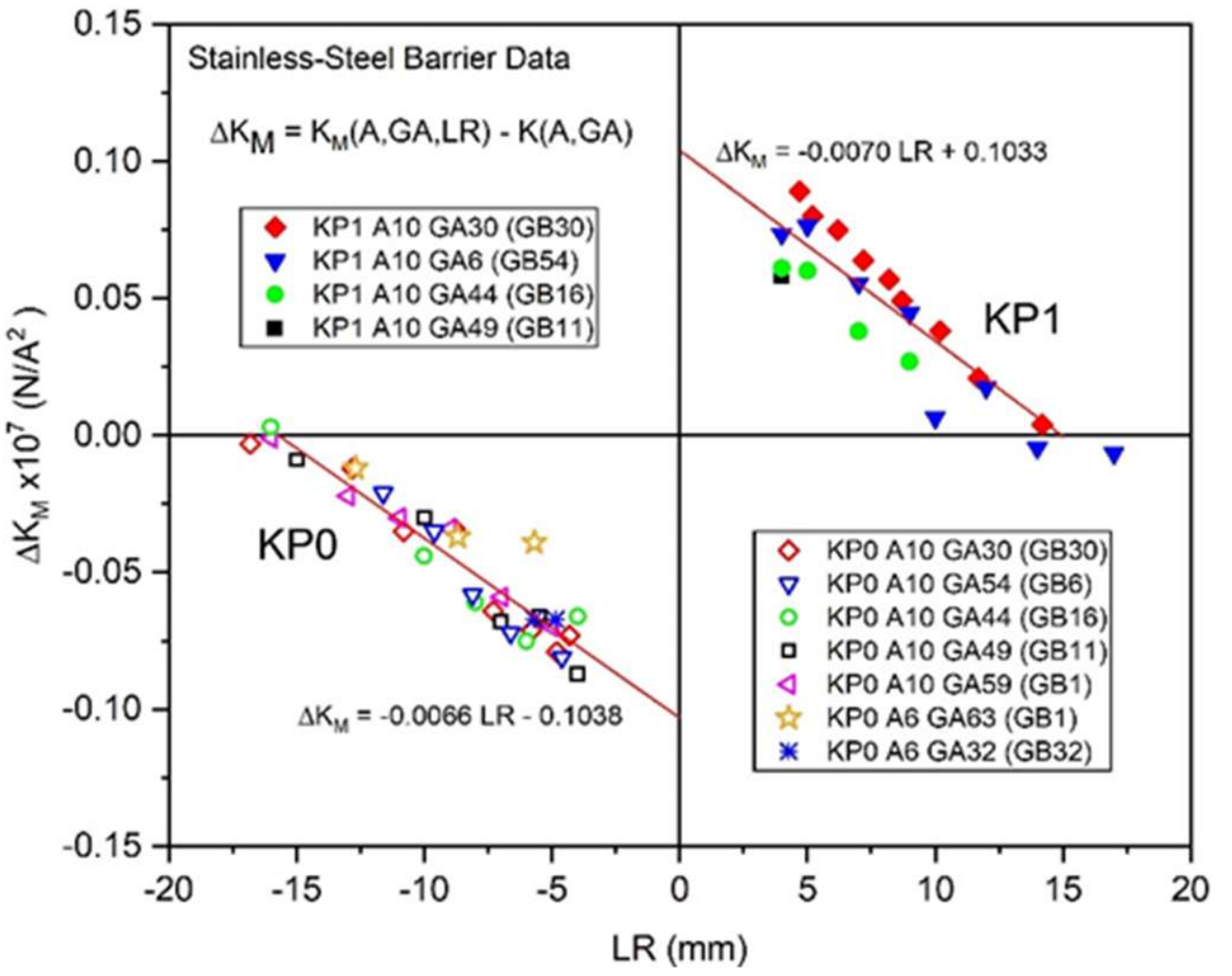

**Fig. 9.** Graph of $\Delta K_M$ vs LR demonstrating that it is purely a function of the lengths of liquid metal column adjacent to its two ends and not a function of A or GA.

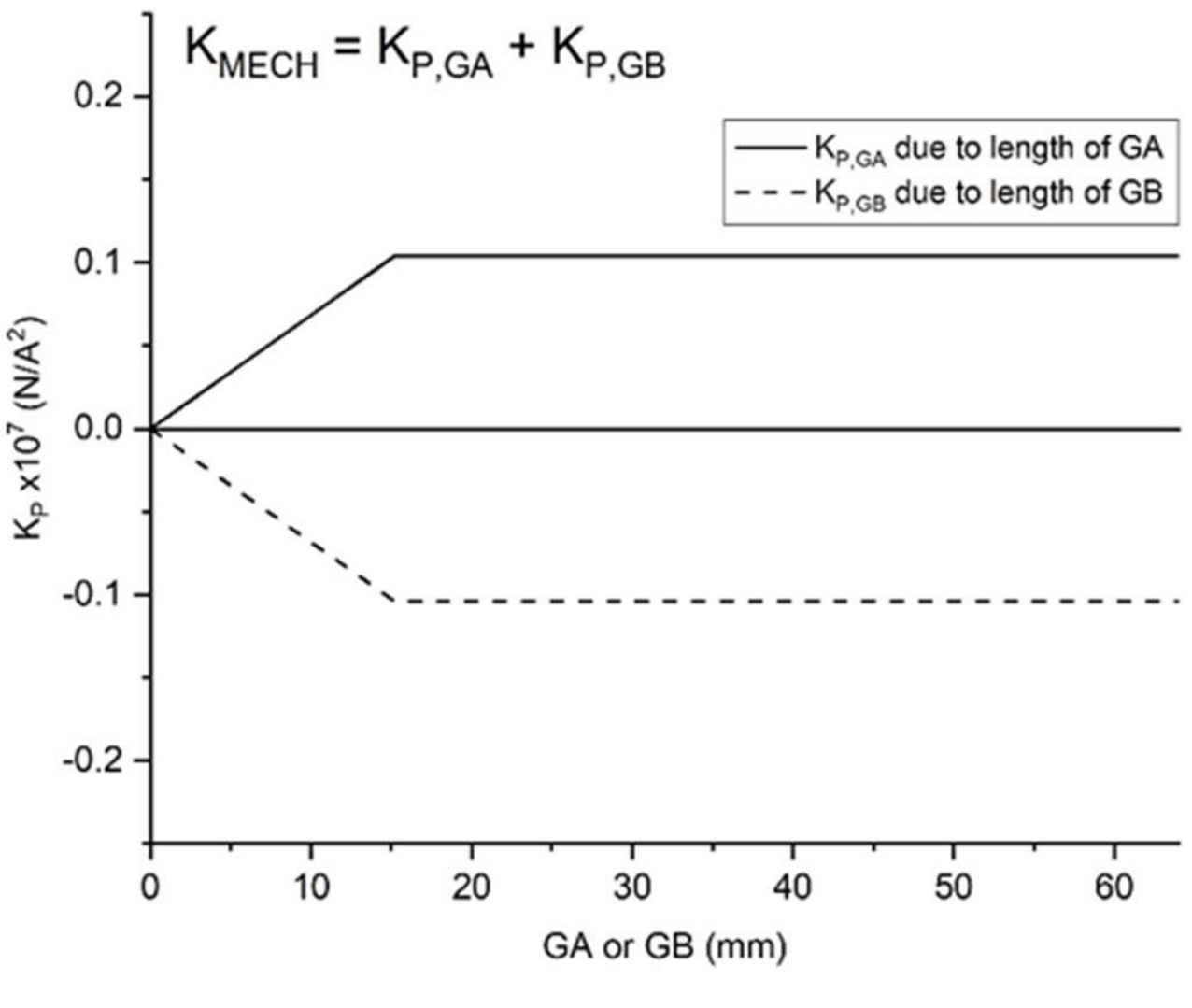

**Fig. 10.** The net mechanical pushing force constant, $K_P$ , on an armature can be estimated by the sum of a positive force due to the liquid in GA and a simultaneous opposing negative force due to the liquid in GB

The slopes of the two straight lines in Fig. 9 are very similar to each other as are the absolute values of the X and Y intercepts. The consistency of this graph therefore indicates that $\Delta K_M$ is a function of LR and not of A or GA. It appears that when the barrier is more than 15 mm from the armature, it makes no perceptible difference to the measured armature force. However, when the barrier is less than 15 mm from the armature, there is progressively less mechanical force applied from that side as the barrier as it is moved toward it, implying a force which decreases with decreasing liquid metal column length. If we extrapolate the data to LR = 0 mm, it yields an estimate of the maximum mechanical column **P**ushing force produced by that side of the trough prior to intervention with the plate to be $K_P = \pm 0.104$. These discoveries invite a simplistic depiction of this pushing force as a function of the unmodified gap lengths, GA and GB as shown in Fig. 10. This graph allows an estimation of the axial mechanical force on the armature from the liquid metal on both of its ends separately which can then be summed to yield an estimate of the net axial mechanical force acting on it. This technique provided a very strong signal and revealed a justifiable estimate of the magnitude and direction of the net mechanical axial force constant, $K_{MECH}$, on the armature for all armature lengths and locations.

*D. EM Force Analysis*

The EM axial force, $K_{EM}$, was deduced by subtracting the estimated mechanical force, $K_{MECH}$ (derived using Fig. 10), from the measured force, K. Consequently, the full data set from Fig. 6 can be re-analysed to produce an estimate of the longitudinal EM force on all three armatures at all locations as depicted in Fig. 11.

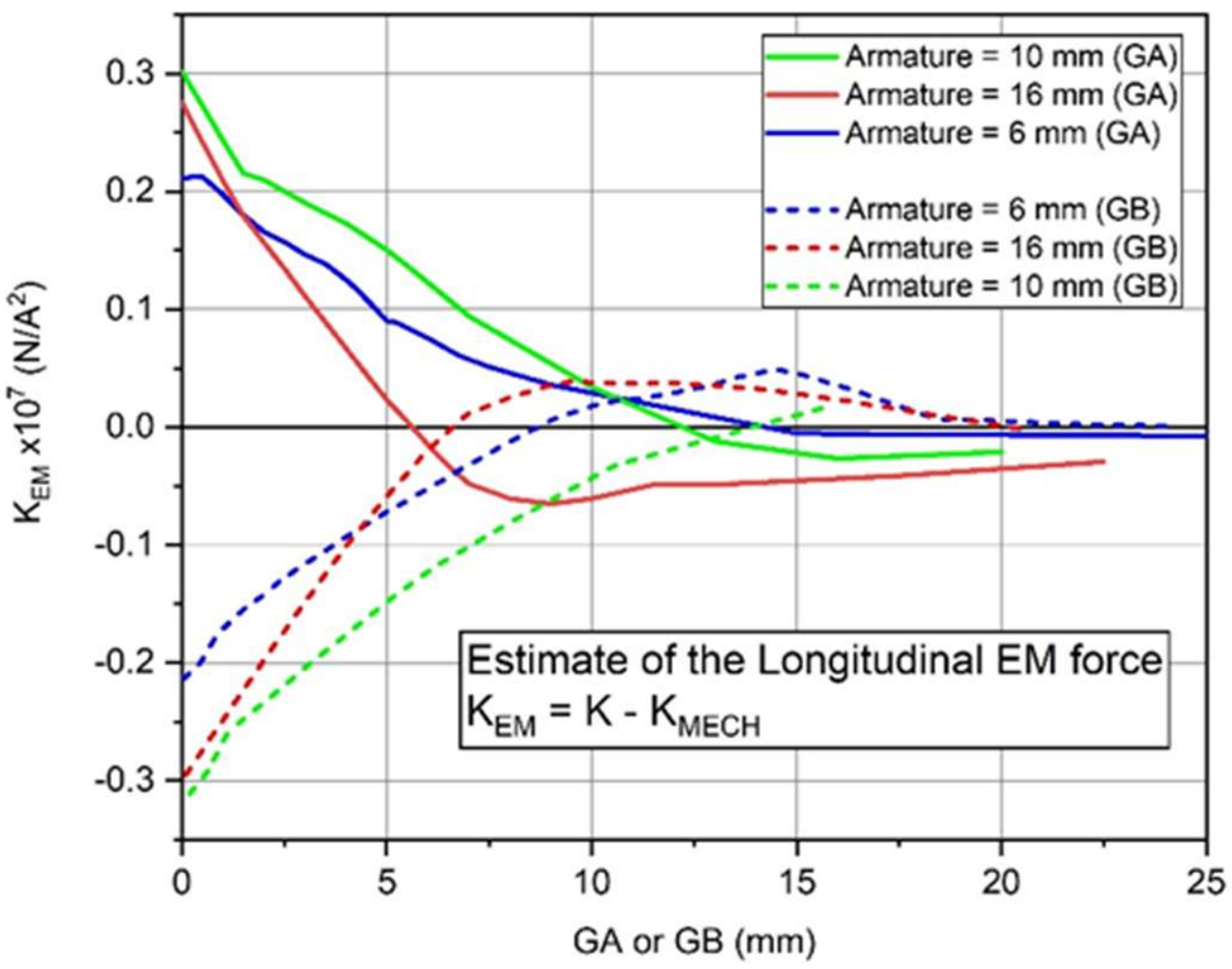

**Fig. 11.** A summary of the estimated longitudinal EM force on the armature for all 3 armatures and all locations.

The fact that for low values of GA or GB, the deduced EM force in Fig. 11 is greater magnitude than the measured force in Fig. 6 occurs because the mechanical force in these regions is in the opposite direction to the EM force. In general, the longitudinal EM force is largest on the longer armatures than on the shorter one. In addition, the EM Force is largest when the armature is nearest to a solid electrode at either of the ends of the trough and pushes the armature away from the electrode. The largest values of the EM force constant, $K_{EM} \sim 0.3$ correspond to 3.8 mN at 355 A DC circuit current. The force decays as the distance from the end electrode increases. In reality, the EM force probably actually tends monotonically to zero at the middle of the trough, however the estimate seems to lose accuracy when the armature is located nearer the centre of the trough. This probably occurs because in this region the measured force is low, and the net mechanical force is calculated by the small difference between two large and possibly slightly inaccurately estimated opposing forces. Hence for the more central armature locations, the EM force estimate is the difference between two small numbers and prone to larger percentage error than when the armature is near the ends of the trough.

Nevertheless, this data implies the existence of a longitudinal EM force on the armature and its magnitude is greater the longer the armature which is to be expected since a larger one contains more atoms, each of which are pushed as a function of the square of the currents which pass near or through them. It also indicates that the EM force on the armature is stronger when it is in closer proximity to a fixed electrode. This finding is more subtle to interpret and requires significantly more experimentation to be sure of the cause. However, a related experiment [5] has already indicated that a variation in mass density within a linear conductor can affect the net longitudinal EM force on internal current elements within it. With copper and Galinstan in the circuit, this mass density variation also occurs in the CRE. However, the most important discovery from the CRE is the demonstration of a steady longitudinal EM force caused by constant DC current, which cannot be predicted by the classical EM force law, namely the Lorentz force.

### *E. Evidence of opposing EM and Mechanical Force*

The expected signature difference between longitudinal EM force and the mechanical pushing force from the liquid metal is the time between the start of the current pulse and the onset of the respective force. On the slow timescale of the CRE, the EM force is expected to start as soon as the current switches on. However, the full mechanical forces in a liquid metal conductor in a trough require the development of height differences and circulation patterns. These were visually observed to take time to develop. It was consequently decided to seek a measurable force demonstration of this differing behaviour. It was anticipated that this might be possible at the armature locations where the net measured force was near zero, but not when the armature was in the middle of the trough. Two such locations for the A = 10 mm armature are anticipated in Fig. 5 and highlighted by stars and described as “sequences used to demonstrate force timing”.

Fig. 12 displays 24 force measurements at these two locations and six different current pulse magnitudes. The current pulses are to scale and shown in red and the force scale is shown on the left. It will be seen that even at the maximum current (355 A), the average measured force was only ~ 0.3 mN. The force measurements at 158.8 A and 194.4 A were consequently only 20% and 30% of that magnitude and are much less distinct. Nevertheless, all these traces reveal an important similarity. At current onset, there is an initial force which reverses in sign 1-1.5 seconds later. On the left-hand side of Fig. 12 when GA = 10.00 mm, the initial force (presumably EM) was positive as shown in Fig. 11, and the later force which is presumably the sum of the EM plus net mechanical force, was negative, agreeing with Fig. 5, and indicating that the net mechanical force, $K_{MECH}$ , was negative in this situation as implied by Fig. 10 when GA = 10.0 and GB = 50.0.

$$K_{MECH} = (0.069) + (-0.104) = -0.035\times10^{-7}\ (N/A^2)$$

In contrast when GB = 12.3, the initial (presumably EM) force was negative, and the later total force (EM plus mechanical) is seen to be positive, again agreeing with Figs. 5, 10 & 11.

This is strong evidence that there are two dominant and often opposed net forces in the CRE, which justifies the analysis technique described in Section IIC. It also further strengthens the claim of the existence of a longitudinal EM force acting directly on the armature.

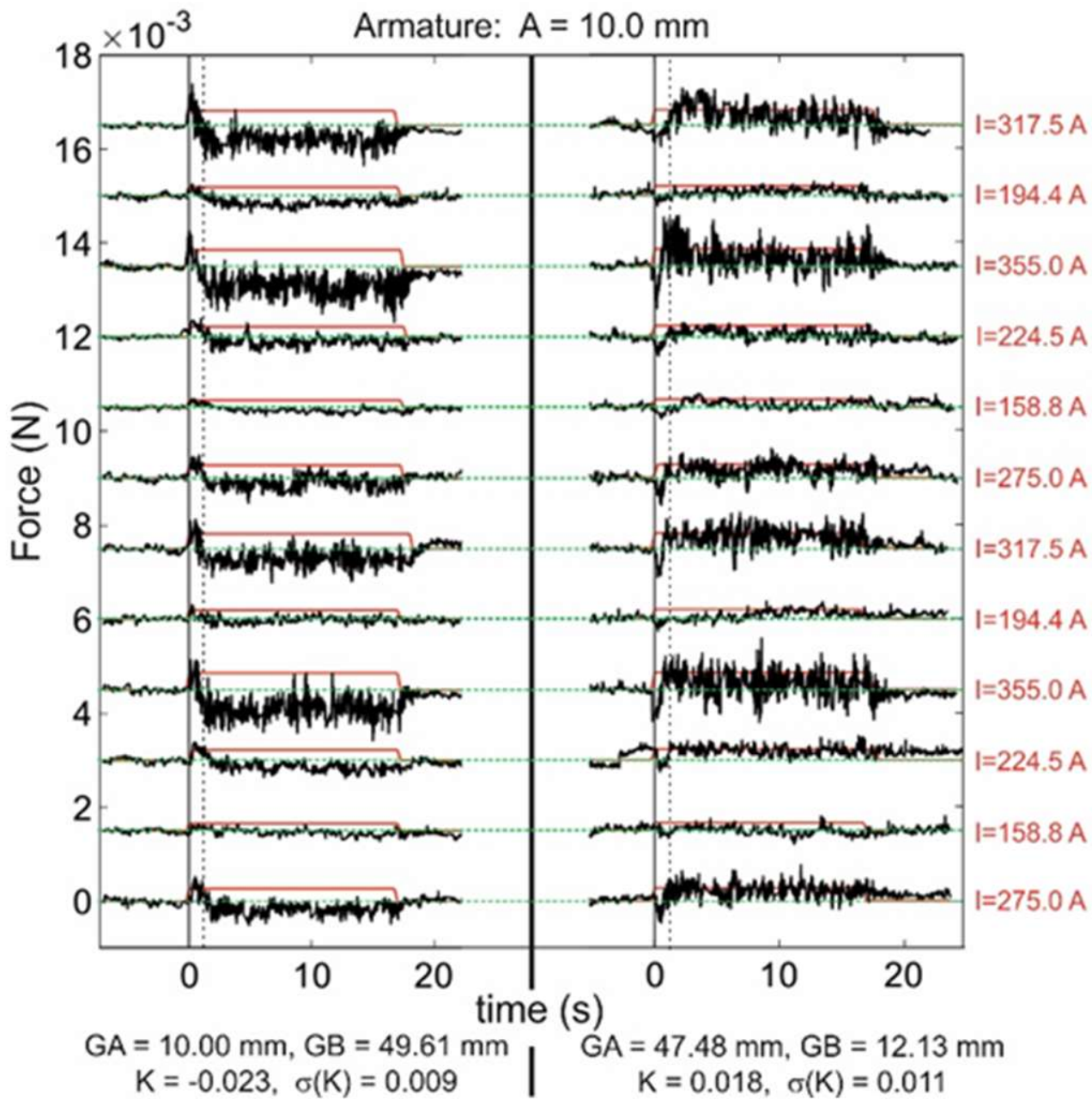

**Fig. 12.** Evidence of the early onset of an EM force on the armature and the later onset of an oppositely directed net EM plus mechanical force in two armature locations at 6 current levels during the experiments with A=10.

## III CONCLUSION

Understanding the measurements of net force on the copper armature immersed in liquid metal in the CRE required a complex analysis of the simultaneous EM and mechanical forces acting on it. It was possible with the CRE to perform a set of measurements that allowed the estimation of the net axial mechanical force acting on the armature. In this way it was possible to deduce the existence and approximate magnitude of longitudinal EM force parallel to the direction of the current, which cannot be predicted by the standard Lorentz force. The strength and direction of the longitudinal EM force was estimated for a range of armature lengths and locations, and in all cases the force was found to be proportional to the square of the circuit current with all other parameters held constant. Although not in common use, there exist historical non-Lorentzian EM force laws, proposed by Ampère [1][2], Weber [7][26] and others [27], which concur with the transverse force prediction of the Lorentz force, but in addition, contain a longitudinal EM force component and therefore suggest a re-inspection as a consequence of the discoveries reported here.

While the CRE provided a qualitative confirmation of longitudinal EM force, it is still not an ideal experiment for isolating and identifying the most accurate form of the EM force law because the liquid metal adds significant complexity to any modelling programme. This will be true for any experiment which relies on plasma or liquid metal as the means of allowing armature displacement while maintaining current continuity. Nevertheless, further research into the longitudinal EM force component is ultimately expected to have a significant impact on physics theory as well as the modelling and design of existing and future pulsed power and high energy plasma applications including fusion, and even lower current density MHD devices.

ACKNOWLEDGEMENT

N. Graneau would like to thank his AWE colleagues, A. Jones and M. Sinclair for their support and very useful discussions throughout the course of this research. He also thanks A. Kharicha at Montan Universität, Leoben for his demonstration of the complexity of liquid metal MHD modelling, and S. Miller and J. Paice at Matrix Precision Engineering Ltd. for their assistance in the design and construction of the CRE experimental apparatus and Figures 1 and 2

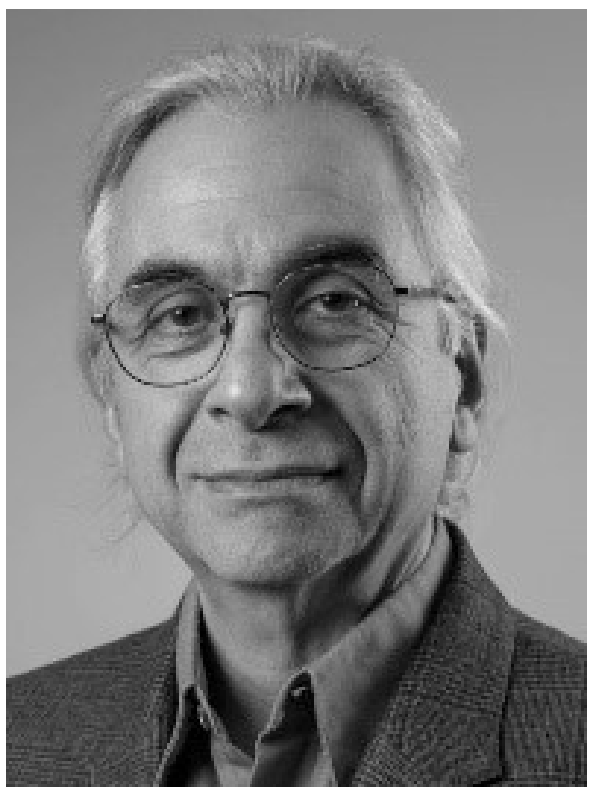

**Neal Graneau** (Member IEEE) was born in London, UK, in 1963. He received the BSc. degree in physics from King's College, London University in 1986 and the MSc. and D.Phil. degrees in plasma physics from the University of Oxford in 1987 and 1992 respectively. His thesis examined ion acceleration mechanisms in the cathode spot of a vacuum arc.

He was a Post-Doc Senior Investigator in the Dept. of Engineering Science at the University of Oxford from 1992-2006 where he led various research projects from the development of novel pulsed power transformers to renewable electricity generation with high current density water arc explosions. Since 2006, he has been a Senior Applied Scientist at AWE Nuclear Security Technologies, leading the development of high current pulsed power machines for dynamic material properties research. A continuous thread throughout his career has been the historical, theoretical and experimental exploration into the fundamental laws and applications of EM force. He has co-authored a book on this subject, *Newtonian Electrodynamics* (World Scientific, 1996). This research feeds into a parallel scientific interest in the philosophical understanding of matter interaction and the cause of inertia which featured in another co-authored book, *In the Grip of the Distant Universe* (World Scientific, 2006).

Dr. Graneau is a fellow of the Institute of Physics